\documentclass[5p]{elsarticle}

\usepackage{natbib}
\usepackage{amssymb,amsmath,xcolor,graphicx,xspace,colortbl,rotating} 
\usepackage{amsmath}  
\usepackage{amssymb,amsmath,xcolor,graphicx,xspace,colortbl,rotating}  
\usepackage{amssymb,amsmath,xcolor,graphicx,xspace,colortbl,textcomp, rotating}  
\graphicspath{{Critical Log Gravity 01_graphics/}{Critical Log Gravity 01_tcache/}{Critical Log Gravity 01_gcache/}}
\DeclareGraphicsExtensions{.pdf,.eps,.ps,.png,.jpg,.jpeg}
\graphicspath{{Critical Log Gravity 01_graphics/}{Critical Log Gravity 01_tcache/}{Critical Log Gravity 01_gcache/}}
\DeclareGraphicsExtensions{.pdf,.eps,.ps,.png,.jpg,.jpeg}
\graphicspath{{Critical_gravity_graphics/}{Critical_gravity_tcache/}{Critical_gravity_gcache/}}
\DeclareGraphicsExtensions{.pdf,.eps,.ps,.png,.jpg,.jpeg}
\graphicspath{{Critical_gravity_graphics/}{Critical_gravity_tcache/}{Critical_gravity_gcache/}}
\DeclareGraphicsExtensions{.pdf,.eps,.ps,.png,.jpg,.jpeg}

\begin{document}

\title{Noether-Wald energy in Critical Gravity}
\author[1]{Giorgos Anastasiou\corref{cor1}}
\ead{georgios.anastasiou@unab.cl}
\author[1]{Rodrigo Olea}
\ead{rodrigo.olea@unab.cl}
\author[1]{David Rivera-Betancour}
\ead{d.riverabe@gmail.com}
\address[1]{Departamento de Ciencias F{\'\i}sicas, Universidad Andres Bello, Sazi{\'e} 2212, Piso 7, Santiago, Chile }

\cortext[cor1]{Corresponding author}

\begin{abstract}
Criticality represents a specific point in the parameter space of a higher-derivative gravity theory, where the linearized field equations become degenerate. In 4D Critical Gravity, the Lagrangian contains a Weyl-squared
term, which does not modify the asymptotic form of the curvature. The Weyl$^{2}$ coupling
is chosen such that it eliminates the massive scalar mode and it renders the
massive spin-2 mode massless. In doing so, the theory turns consistent around
the critical point.\newline
Here, we employ the Noether-Wald method to derive the conserved quantities for the action of
Critical Gravity. It is manifest from this energy definition that, at the critical point, the mass is identically zero
for Einstein spacetimes, what is a defining property of the theory. As the entropy is obtained
from the Noether-Wald charges at the horizon, it is evident that it also vanishes for any Einstein black hole.

\end{abstract}

\maketitle

\section{Introduction}

General Relativity (GR) is a successful theory of gravity at a classical level but it lacks of consistency
in a quantum regime because it is not renormalizable. On the other hand, in the low energy
limit of String Theory, which should be finite to all orders, there appear contributions that are quadratic
in the curvature.
As a consequence, higher curvature extensions of Einstein gravity are expected to give rise to a
gravity theory with a better ultraviolet behavior.
Early work on the subject has suggested that this class of theories should be renormalizable \citep{Stelle:1976gc}.

Lower-dimensional examples have been extensively studied in recent literature. They are regarded as
insightful toy models which capture essential features of $4D$ gravity. One of them is New Massive
Gravity (NMG) \citep{Bergshoeff:2009hq}, a parity-even three-dimensional theory which describes
two propagating massive spin-2 modes, in contrast to $3D$ Einstein gravity which is topological.
Picking up the conventional sign of the Einstein-Hilbert action, the energy of the massive excitations
is negative (ghost modes), while the mass of the Banados-Teitelboim-Zanelli (BTZ) black hole is positive.
Clearly, this inconsistency persists even if one reverses the sign of the kinetic term. A physically
reasonable theory arises at a specific point of parametric space, where the massive spin-2 field
turns massless \citep{Liu:2009bk}. At this particular point, both the energy of the graviton and the mass
of the BTZ black hole vanish identically \citep{Clement:2009gq}. Furthermore, both central charges turn into
zero, what leads to a vanishing entropy \citep{Clement:2009gq}. Another feature of the theory is the presence
of new modes with logarithmic behavior at the critical point \citep{Liu:2009kc}. These modes are
eliminated when standard Brown-Henneaux boundary conditions are considered. Relaxing the asymptotic
conditions to include log terms switches on new holographic sources at the boundary \citep{Grumiller:2008es}.

Another theory in three dimensions sharing similar features with NMG is Topologically Massive Gravity(TMG) \citep{Deser:2002iw}. The corresponding critical point defines the concept of Chiral Gravity. However, in this case, the central charges are different from each other due to a parity-violating term in the action. As a consequence, neither mass nor entropy vanish for BTZ black holes at the chiral point.

The generalization of the concept of criticality, present in these models, to four dimensions is given
by theories which include quadratic terms in the curvature with particular couplings on top of the Einstein-Hilbert action.
The most general form of a gravity action with quadratic-curvature corrections  in $4D$ is given by

\begin{equation}  \label{quadratic_gravity}
I=\frac{1}{16\pi G}\int \limits_{M} d^{4}x\sqrt{-g}\left(R-2\Lambda+\alpha
R_{\mu\nu}R^{\mu\nu}+\beta R^{2} \right)\,,
\end{equation}

\noindent
where $\alpha$ and $\beta$ are arbitrary couplings, and $\Lambda=-3/\ell^{2}$ is the cosmological constant
in terms of the AdS radius $\ell$.
The Riemann-squared term is not present, as it can be always traded off by the Gauss-Bonnet (GB)
invariant plus the curvature-squared terms present in the action (\ref{quadratic_gravity}). The
GB term does not affect the field equations in the bulk but it does modify the boundary dynamics.

This class of theories leads to equations of motion (EOM) with up to four derivatives in the metric.
Generically, they describe modes that represent a massless spin-2 graviton, a massive spin-2 field
and a massive scalar.
For a quadratic-curvature gravity theory with arbitrary coupling constants, perturbations around a
given background would give rise to ghosts. The problem with the sign of the energy of these modes
can be circumvented by a sign flip of the constant in front of Einstein kinetic term. On the other
hand, Einstein black holes are solutions to the theory defined by Eq.(\ref{quadratic_gravity}).
Therefore, the change in the sign mentioned above would lead to a negative mass for Schwarzschild-AdS
black hole. Needless to say, this picture is clearly unphysical as the energy of the perturbations
around a background and the mass of a black hole carry opposite signs.

In view of this general obstruction to obtain a four-dimensional gravity
theory which is free of the inconsistencies discussed above, it was quite surprising when the authors of Ref.\citep{Lu:2011zk}
pointed out the fact that, for the particular couplings $\alpha=-3\beta$ and $\beta=-1/2\Lambda$, the massive scalar
is eliminated and the massive spin-2 mode turns massless.
This choice renders the theory physically sensible around the critical point. This fact is confirmed
by using the Ostrogradsky method for Lagrangians with derivatives of higher order: the energy for the
massive mode vanishes for the critical value of the couplings.
From the point of view of the energy of the black holes of the theory, one can use the Abbott-Deser-Tekin (ADT)
formula  \citep{Deser:2002jk,Deser:2002rt} to evaluate the mass of Schwarzschild-AdS solution, what results in

\begin{equation}  \label{masa}
M=m\left(1+2\Lambda(\alpha+4\beta)\right)\,,
\end{equation}

\noindent
where $m$ is the mass parameter in the solution. \newline
The general formula Eq.(\ref{masa}), makes evident that, for the critical condition mentioned
above, the mass for Schwarzschild-AdS black hole vanishes.

In the present work, as an alternative to Deser-Tekin procedure, we employ Noether-Wald method \citep{Iyer:1994ys,Iyer:1995kg}
to compute the charges in Critical Gravity.
This full (non-linearized) expression derived in this way has a remarkable property: the energy of any
Einstein space is identically zero, as long anticipated in Ref.\citep{Porrati:2011ku}.

\section{Deser-Tekin energy in $4D$ quadratic-curvature gravity} \label{Deser-Tekin}

As  mentioned in the previous section, in Refs.\citep{Deser:2002jk,Deser:2002rt}, the authors provide
a generic definition of energy for an arbitrary curvature-squared gravity theory.
That definition of the energy is obtained as an extension of the Abbott-Deser method
\citep{Abbott:1981ff}.

In order to obtain the ADT mass for a general asymptotically AdS (AAdS) solution, we need
to write down the metric of the spacetime in the form of $g_{\mu \nu }=\bar{g}_{\mu \nu }+h_{\mu \nu }$,
where $\bar{g}_{\mu \nu }$ is the metric of the background and $h_{\mu \nu }$ is the perturbation tensor.
Such construction leaves the first-order variation of field equations as

\begin{align}
&\delta\left( G_{\mu \nu }+E_{\mu \nu }\right) =\left[ 1+2\Lambda \left(
\alpha +4\beta \right) \right] G_{\mu \nu }^{L} \notag \\
&+\alpha \left[ \left( \bar{%
\square}-\frac{2\Lambda }{3}\right) G_{\mu \nu }^{L}-\frac{2\Lambda }{3}R^{L}%
\bar{g}_{\mu \nu }\right] + \notag \\
&+\left( \alpha +2\beta \right) \left[ -\bar{\nabla}_{\mu }\bar{\nabla}_{\nu
}+\bar{g}_{\mu \nu }\bar{\square}+\Lambda \bar{g}_{\mu \nu }\right] R^{L}\,, \label{EOM1}
\end{align}%

\noindent
where $G_{\mu \nu }^{L}$ and $R^{L}$ are the linearized expression of Einstein
tensor and Ricci scalar, respectively. The tensor
$E_{\mu \nu }$ is the contribution of fourth order in the derivatives to
the field equations. The equation (\ref{EOM1}) has to be equal to an
effective energy-momentum tensor $T_{\mu \nu }$, which is covariantly conserved. One can write a conserved current,
for a set of Killing fields $\{\bar{\xi} ^{\mu }\}$ that represents the isometries of the background

\begin{equation}
J_{ADT}^{\mu } = 8 \pi G T^{\mu \nu }\bar{\xi}_{\nu }\,.
\end{equation}%

\noindent
In order to evaluate the mass of a gravitational object, the Killing vector needs to be timelike, at least, at infinity. \newline
Whenever there is a current which is conserved, one is able to write down $J^{\mu }$
as the divergence of a 2-form prepotential, i.e.,

\begin{equation} \label{Jdsth}
J_{ADT}^{\mu }=\nabla _{\nu }\mathcal{F}^{\mu \nu }\,.
\end{equation}

\noindent
One can consider a spacetime foliated by a normal (radial) direction $z$

\begin{equation} \label{GaussC}
ds^{2}=N^{2}(z)dz^{2}+h_{ij}(z,x)dx^{i}dx^{j}\,,
\end{equation}

\noindent
where $h_{ij}(z,x)$ is the induced metric on $\partial M$, and its radial evolution is defined by the unit
vector $n_{\nu}=N(z)\delta^{z}_{\nu}$.

In this coordinate frame, the conserved charge can be expressed as an integral on the co-dimension two surface $\Sigma$

\begin{equation} \label{QintSigma}
Q_{ADT}^{\mu }\lbrack \bar{\xi} \rbrack = \int \limits_{\Sigma}dS_{\nu}\mathcal{F}%
^{\mu \nu}\,.
\end{equation}

\noindent
Here, $dS_{\nu}= d^{2}x \, \sqrt{-h}\,n_{\nu}$ is a surface normal vector that defines the integration
for a fixed time and radius.
For the case of curvature-squared gravity in four dimensions, the conserved quantity adopts the form
\footnote{For a generalized ADT procedure see, e.g., Ref.\citep{Peng:2014gha}}

\begin{align}
&8\pi G\,Q_{ADT}^{\mu }\lbrack \bar{\xi} \rbrack =\left[ 1+2\Lambda \left( \alpha +4\beta \right) %
\right] \int \limits_{\partial M} d^{3}x\,G_{L}^{\mu \lambda }\bar{\xi}_{\lambda } \notag \\
&+\left( \alpha +2\beta
\right) \int \limits_{\Sigma} dS_{\nu}\left( 2\bar{\xi}^{\,[\mu }\bar{\nabla}^{\nu]}R^{L}+R^{L}\bar{%
\nabla}^{\mu }\bar{\xi}^{\nu}\right)  \notag \\
&-\alpha \int \limits_{\Sigma} dS_{\nu}\left( 2\bar{\xi}_{\lambda }\bar{\nabla}^{[\mu
}G_{L}^{\nu]\lambda }+2G_{L}^{\lambda [\mu }\bar{\nabla}^{\nu]}\bar{\xi}%
_{\lambda }\right)\,. \label{QADT}
\end{align}%

\section{Critical Gravity}

In Ref. \citep{Lu:2011zk}, the energy of the graviton modes in quadratic-curvature gravity  was studied.
These excitations come from the linearized EOM (\ref{EOM1}). The choice $\alpha =-3\beta $ leads to a traceless
perturbation ($h=0$) which eliminates the massive scalar mode. Consequently, the equation for the propagating mode takes
the form

\begin{equation}
\left(\bar {\Box} -\frac{2\Lambda }{3}\right) \left(\bar{\Box}  -\frac{2\Lambda }{3}-%
\frac{2\Lambda \beta +1}{3\beta }\right) h_{\mu \nu }=0.
\end{equation}

\noindent
The first factor of the equation describes the propagation of
a massless graviton in an AdS background while the second one represents a
massive spin-2 field.

It is clear that the latter becomes massless by imposing the
critical value $\beta =-1/2\Lambda $. This particular coupling produces the
fourth order equation

\begin{equation}
\left( \bar{\Box} -\frac{2\Lambda }{3}\right) ^{2}h_{\mu \nu }=0 \,,
\end{equation}

\noindent
which reflects the appearance of both massless and logarithmic modes \citep{Lu:2011zk}.

In order to obtain the energy of the excitations, the authors in Ref. \citep{Li:2008dq}
followed a Hamiltonian approach.  For an unrestricted value of$\beta $, the action  up to quadratic order in  $h^{\mu \nu }$ is

\begin{align}
&I=-\frac{1}{16\pi G}\int \limits_{M} d^{4}x\sqrt{-g}\left[ \frac{1}{2}\left( 1+6\beta
\Lambda \right) \bar{\nabla}^{\lambda }h^{\mu \nu }\bar{\nabla}_{\lambda
}h_{\mu \nu } \right. \notag\\
& \left. +\frac{3}{2}\beta \bar{\Box}h^{\mu \nu }\bar{\Box}h_{\mu \nu }+%
\frac{\Lambda }{3}\left( 1+4\beta \Lambda \right) h^{\mu \nu }h_{\mu \nu }%
\right] \,.
\end{align}

\noindent
Using the Ostrogradsky method for higher-derivative Lagrangians, one
obtains the following conjugate momenta

\begin{align}
&\pi_{(1)}^{\mu \nu } &=&\frac{1}{16\pi G}\sqrt{-g}\bar{\nabla}^{0}\left[ \left( 1+6\beta
\Lambda \right) h^{\mu \nu }-3\beta \bar{\Box}h^{\mu \nu }\right] \,, \\
&\pi_{(2)}^{\mu \nu } &=& \frac{3\beta }{16\pi G}\sqrt{-g}\bar{g}^{00}\bar{\Box}h^{\mu \nu
} \,.
\end{align}

\noindent
Due to the fact that the Lagrangian is time independent, the
Hamiltonian can be written as its time average, that is

\begin{align}
&H=\frac{1}{16\pi GT}\int \limits_{M} d^{4}x\sqrt{-g}\left[ \left( 1+6\beta \Lambda
\right) \bar{\nabla}^{0}h^{\mu \nu }\dot{h}_{\mu \nu } \right. \notag\\
&\left.-6\beta \left( \frac{%
\partial }{\partial t}\left( \bar{\Box}h^{\mu \nu }\right) \right) \bar{%
\nabla}^{0}h_{\mu \nu }\right] -\frac{1}{T}I \,.
\end{align}

\noindent
Evaluating for the case of  massless and massive propagating modes, one
obtains the following expressions for the corresponding on-shell energies

\begin{align}
&E_{(m)} = - \frac{1}{16\pi GT}\left( 1+2\beta \Lambda \right) \int \limits_{M} d^{4}x\bar{%
\nabla}^{0}h_{m}^{\mu \nu }\dot{h}_{\mu \nu }^{m}\,, \label{Em} \\
&E_{(M)} = \frac{1}{16\pi GT}\left( 1+2\beta \Lambda \right) \int \limits_{M} d^{4}x\bar{%
\nabla}^{0}h_{M}^{\mu \nu }\dot{h}_{\mu \nu }^{M}\,, \label{EM}
\end{align}

\noindent
where the subscripts $m$ and $M$ stand for massless graviton and
massive spin-2 field, respectively.

In a gravity theory with quadratic terms in the curvature, where the couplings are related as $\alpha =-3\beta $,
there is only a specific value of $\beta$ that kills the negative energy states. More specifically, from Eqs. (\ref{Em},\ref{EM})
it is shown that for $\beta = -1/2\Lambda$, the energy of both the massless and the massive modes is zero.
Hence, all the ghosts disappear leading to a consistent theory of gravity.

Therefore, the action of Critical Gravity reads

\begin{equation} \label{CritG_action}
I_{critical}=\frac{1}{16\pi G}\int \limits_{M} d^4x\sqrt{-g} \left[\left(R+\frac{6}{\ell^{2}}\right)-\frac{%
\ell^{2}}{2}\left(R_{\mu\nu}R^{\mu\nu}-\frac{1}{3}R^2\right)\right] \,.
\end{equation}

\noindent
On the other hand, the generic expression for the energy of the black holes in this gravity theory is given
by Eq.(\ref{QADT}). For any static black hole, the only nonvanishing contribution comes from the first
term on the right hand side of Eq.(\ref{QADT}). In particular, for a Schwarzschild-AdS black hole, the
ADT charge leads to the result in Eq.(\ref{masa}). Is is easy to notice that, for the particular value
of the couplings which define Critical Gravity ($\alpha =-3\beta $, $\beta=-1/2\Lambda $), the mass of
the black hole vanish.

In what follows, we provide an alternative formula of conserved charges in Critical Gravity, which makes
manifest the fact that the energy for Einstein black holes is identically zero.

\section{Noether-Wald charges in Critical Gravity}

A general prescription to define conserved charges in an arbitrary theory of gravity was given
in Refs.\citep{Iyer:1994ys,Iyer:1995kg,Wald:1999wa}. For the purpose of the discussion below, we will restrict
ourselves to the case where Lagrangian density is a functional only of the metric and the curvature,
${\cal{L}}(g_{\mu\nu},R_{\mu\nu\alpha\beta})$.
For a given set of Killing vectors $\{\xi^{\mu}\}$, the Noether current is written down as

\begin{equation}  \label{corriente}
\sqrt{-g}J^{\mu}=\Theta^{\mu}\left(\delta_{\xi}g\right)+\Theta^{\mu}\left(%
\delta_\xi\Gamma\right)+\sqrt{-g}\cal{L}\xi^{\mu}\,.
\end{equation}

\noindent
For simplicity, we assume that the surface term $\Theta^{\mu}$ is separable into a part that
contains variations of the Christoffel symbol and another part that contains variations of the metric. As we
are interested in diffeomorphic charges for gravity, all the variations are replaced by a Lie derivative
along the vector $\{\xi^{\mu}\}$.

Using the Killing equation, $\delta_\xi
g_{\mu\nu}=\nabla_\mu\xi_\nu+\nabla_\nu\xi_\nu=0$, one can notice that first term in Eq.(\ref{corriente})
vanishes. The same relation, this time for the Lie derivative of the Christoffel connection, would produce a combination of double covariant derivatives and curvatures. This casts the current, for a generic gravity theory, in the form

\begin{equation}  \label{corriente_2}
J^{\mu}=2E^{\mu\nu}_{\alpha\beta}\left(\nabla_\nu\nabla^\alpha\xi^%
\beta+R^{\alpha\beta}_{\nu\sigma}\xi^\sigma\right)+\xi^\mu \cal{L}\,.
\end{equation}

\noindent
Here, the tensor $E^{\mu\nu}_{\alpha\beta}$ is the functional derivative of ${\cal{L}}$ with respect
to the spacetime Riemann tensor $R^{\mu\nu}_{\alpha\beta}$, that is,

\begin{equation}  \label{tensorE}
E^{\mu\nu}_{\alpha\beta}=\frac{\delta \cal{L}}{\delta R^{\alpha\beta}_{\mu\nu}}%
\,.
\end{equation}

\noindent
It can be shown, by means of the general form of the field equations for these class of gravity theories,
that the last two terms on the right hand side of (\ref{corriente_2}) form the EOM contracted
with the Killing field.

Thus, on-shell, the first term on the right side of (\ref{corriente_2}) is the only nonvanishing part.

As the tensor $E^{\mu\nu}_{\alpha\beta}$ satisfies Bianchi identity, the conserved current turns into
a total derivative

\begin{equation} \label{corriente_3}
J^{\mu}=2\nabla_{\nu}\left(E^{\mu\nu}_{\alpha\beta}\nabla^{\alpha}%
\xi^{\beta}\right)\,.
\end{equation}

\noindent
As the Noether current $J^{\mu}$ can be written as $J^{\mu}=\nabla_{\nu}q^{\mu\nu}$, the conserved charge is expressed as an
integral on the co-dimension two surface $\Sigma$

\begin{equation}  \label{carga}
Q^{\mu}[\xi]=\int \limits_{\Sigma}dS_{\nu}q^{\mu\nu}
\end{equation}

\noindent
as mentioned previously in Section \ref{Deser-Tekin}.
Finally the conserved charge is written as

\begin{equation}  \label{carga_2}
Q^{\mu}[\xi]=2\int \limits_{\Sigma}\,dS_{\nu}E^{\mu\nu}_{\alpha%
\beta}\nabla^{\alpha}\xi^{\beta}.
\end{equation}

\noindent
An alternative form for the action of Critical Gravity considers the
difference between Weyl$^{2}$ and the GB term ${\cal{E}}_{4}$, as the GB invariant
term does not alter the bulk dynamics \citep{Miskovic:2014zja}

\begin{align}  \label{CritG_actionA}
I_{critical} &=\frac{1}{16\pi G}\int \limits_{M} d^{4}x\sqrt{-g}\left[\left(R+\frac{6}{\ell^{2}}\right) \right. \notag\\
& \left. +\frac{\ell^2}{4}\left({\cal{E}}_{4}-W^{\mu\nu\alpha\beta}W_{\alpha%
\beta\mu\nu}\right)\right]\,.
\end{align}

\noindent
We can split the action in two parts:  the first one is the MacDowell-Mansouri action, $I_{MM}$, which is given by the
Einstein-Hilbert plus GB terms, the latter with a fixed coupling \citep{MacDowell:1977jt}. In Einstein gravity, this corresponds to
a built-in renormalized AdS action \citep{Miskovic:2009bm}. The second part is minus  the action of Conformal Gravity $I_{CG}$.

Using the Noether-Wald formula for the current (\ref{corriente_3}) for the first part, the functional derivative with respect to the Riemann tensor of the Lagrangian in $I_{MM}$ produces

\begin{equation}  \label{E_MM}
E_{\alpha\beta}^{\mu\nu}=\frac{\ell^{2}}{128\pi G}%
\delta^{[\mu\nu\sigma\lambda]}_{[\alpha\beta\gamma\delta]}\left(R^{\gamma%
\delta}_{\sigma\lambda}+\frac{1}{\ell^{2}}\delta^{[\rho\delta]}_{[\gamma%
\lambda]}\right)\,,
\end{equation}

\noindent
whereas, for the Conformal Gravity part $I_{CG}$, we get

\begin{equation}   \label{E_CG}
\tilde{E}_{\alpha\beta}^{\mu\nu}=-\frac{\ell^{2}}{128\pi G}%
\delta^{[\mu\nu\sigma\lambda]}_{[\alpha\beta\gamma\delta]}W^{\gamma\delta}_{%
\sigma\lambda}
\end{equation}

\noindent
Using the Noether-Wald formula (\ref{carga_2}), the total charge for the
theory

\begin{equation}   \label{total_charge}
Q^{\mu}[\xi]=\frac{\ell^2}{64\pi G}\int \limits_{\Sigma} \,dS_\nu
\delta^{[\mu\nu\sigma\lambda]}_{[\alpha\beta\gamma\delta]}\nabla^{\alpha}%
\xi^{\beta}\left[\left(R^{\gamma\delta}_{\sigma\lambda}+\frac{1}{\ell^{2}}%
\delta^{[\gamma\delta]}_{[\sigma\lambda]}\right)-W^{\gamma\delta}_{\sigma%
\lambda}\right].
\end{equation}

\noindent
By definition, the Weyl tensor is

\begin{equation}  \label{weyl}
W^{\gamma\delta}_{\sigma\lambda}=R^{\gamma\delta}_{\sigma\lambda}-\frac{1}{2}%
\left(R^{\gamma}_{\sigma}\delta^{\delta}_{\lambda}-R^{\delta}_{\sigma}\delta^{\gamma}_{\lambda}-
R^{\gamma}_{\lambda}\delta^{\delta}_{\sigma}+R^{\delta}_{\lambda}\delta^{\gamma}_{\sigma}\right)+\frac{1}{6}R\delta^{[\gamma\delta]}_{[%
\sigma\lambda]}.
\end{equation}

\noindent
For Einstein spaces, $R_{\mu\nu}=- (3/\ell^2) g_{\mu\nu}$, the Weyl tensor
adopts the particular form

\begin{equation}  \label{AdS_curvature}
W^{\gamma\delta}_{(E)\sigma\lambda}=R^{\gamma\delta}_{\sigma\lambda}+\frac{1}{%
\ell^{2}}\delta^{[\gamma\delta]}_{[\sigma\lambda]}\,,
\end{equation}

\noindent
where the right hand side, is referred to as AdS curvature \footnote{The field strength for the AdS group
also contains the torsion along the generators of AdS translations in Riemann-Cartan theory. For Riemannian
geometry, Eq.(\ref{AdS_curvature}) is the only nonvanishing part of the curvature of the AdS group.}
Using the above fact, the conserved quantity in Critical Gravity is
identically zero for Einstein spaces.

\section{Electric part of the Weyl tensor and Einstein modes in Conformal Gravity} \label{electric}

CG in four dimensions is invariant under local Weyl rescalings of the metric ($ g_{\mu \nu} \rightarrow
\tilde{g}_{\mu \nu} = e^{2 \omega} g_{\mu \nu}$). Solutions to CG are Bach-flat geometries, which
include Einstein spacetimes.

From a holographic viewpoint, asymptotically AdS space in CG are endowed with new sources
at the conformal boundary. Indeed, we can set any AAdS spacetime in Fefferman-Graham (FG) form of the metric

\begin{equation}
ds^{2} = \frac{\ell^{2}}{z^{2}} dz^{2} + \frac{1}{z^{2}} g_{i j}(z,x)dx^{i}dx^{j}\,,
\label{FGz}
\end{equation}

\noindent
where the metric $g_{i j}(z,x)$ is expanded as a power series around the boundary $z = 0$, i.e.,

\begin{equation}
g_{i j}(z,x)=g_{\left (0\right ) i j}+ z g_{\left (1\right ) i j}  + z^{2} g_{\left (2\right ) i j} + z ^{3}  g_{\left (3\right ) i j} + ... \,.
\end{equation}

\noindent
Here, the ellipsis denotes higher-order terms which do not enter into the holographic description of $4D$
AAdS spaces.

The presence of the term $ z g_{\left (1\right ) i j}$ reflects the fact the space contains a non-Einstein part. By
demanding the vanishing of the linear term on $z$, one recovers the Einstein branch, with only even powers
of $z$ in the expansion. This is achieved by imposing a Neumann boundary condition on the metric, $\partial_{z} g \mid_{z=0} = 0$ \citep{Maldacena:2011mk}.

On the other hand, the Noether-Wald charge for Conformal Gravity is proportional to the Weyl tensor, as shown
by Eq.(\ref{E_CG}). However, it is not obvious whether, for Einstein spaces, the holographic modes of CG at the boundary are contained
in the electric part of the Weyl tensor

\begin{equation}
E^{i}_{j} = W^{i\mu}_{j\nu}n_{\mu}n^{\nu}=W^{iz}_{jz}\,,
\end{equation}

\noindent
as it is the case in Einstein gravity.

As Einstein spaces are solutions of the EOM of CG in the bulk, we restrict the discussion to the surface term in the variation of $I_{CG}$, that is,

\begin{align}
\delta I_{CG} &= \frac{\ell^{2}}{64\pi G} \int \limits _{\partial M} d^{3}x \sqrt{-h} \delta^{[\mu_1\mu_2\mu_3\mu_4]}_{[\nu_1\nu_2\nu_3\nu_4]} \left[n_{\mu_1} \delta\Gamma^{\nu_1}_{\beta \mu_2} g^{\nu_2 \beta}  W^{\nu_3 \nu_4}_{(E)\mu_3 \mu_4} \right. \notag \\
& \left. + n^{\nu_1} \nabla_{\mu_1} W^{\nu_2 \nu_3}_{(E)\mu_2 \mu_3} \left(g^{-1}\delta g\right)^{\nu_4}_{\mu_4}\right]\,,
\label{conformal action}
\end{align}

\noindent
where $W_{(E)}$ is the Einstein part of the Weyl tensor (\ref{AdS_curvature}).

The second term in the above relation can be eliminated using the Bianchi identity of second kind.
A projection of all indices to the boundary can be performed by taking the explicit form of the normal
vector $n_{\mu}$ in Gaussian coordinates. Then, the surface term takes the form

\begin{align}
\delta I_{CG} &= \frac{\ell^{2}}{64\pi G} \int \limits _{\partial M} d^{3}x \sqrt{-h} \delta^{[i_1i_2i_3]}_{[j_1j_2j_3]} N(z) \left[g^{j_1\ell} \delta\Gamma^{z}_{i_1\ell} W^{j_2j_3}_{(E)i_2i_3} \right. \notag \\
& \left. - g^{z z}\delta\Gamma^{j_1}_{i_1 z} W^{j_2j_3}_{(E)i_2i_3}  - 2 g^{j_2 \ell} W^{j_3 z}_{(E)i_2 i_3} \delta\Gamma^{j_1}_{\ell i_1}\right]\,.
\label{variacion0}
\end{align}

\noindent
In Gauss normal frame (\ref{GaussC}), the relevant components of the Christoffel symbol are

\begin{align}
\Gamma^{z}_{ij} &= \frac{1}{N} K_{ij} \,, \nonumber \\
\Gamma^{i}_{z j} &= - N K^{i}_{j} \,, \nonumber \\
\Gamma^{i}_{jk}\left(g\right) &= \Gamma^{i}_{jk}\left(h\right) \,, \label{gammas}
\end{align}

\noindent
where $K_{ij}=-\frac{1}{2N}\partial_z h_{ij}$ is the extrinsic curvature at $\partial M$.
Equipped with this result, the variation of the action is written as

\begin{align}
\delta I_{CG} &= \frac{\ell^{2}}{64\pi G} \int \limits _{\partial M} d^{3}x \sqrt{-h} \delta^{[i_1i_2i_3]}_{[j_1j_2j_3]} \left[2 W^{j_2j_3}_{i_2i_3} \delta K^{j_1}_{i_1} \right. \notag \\
& \left. + \left(h^{-1}\delta h\right)^{j_1}_{\ell}K^{\ell}_{i_1} W^{j_2j_3}_{i_2i_3}  - 2N \delta\Gamma^{j_1}_{i_1 \ell} h^{\ell j_2}W^{j_3 z}_{i_2i_3}\right]\,,
\label{varCG2}
\end{align}

\noindent
after some algebraic manipulation and index relabeling.

The rest of the proof relies on a power-counting argument in the radial coordinate $z$. In order to do so,
it is required to expand the tensorial quantities which appear at the surface term.

First, we consider the FG expansion for Einstein spacetimes, where $N(z) = \ell/z$ and $h_{ij}(z,x) = g_{ij}(z,x)/z^{2}$
with the metric at the conformal boundary given by

\begin{equation}
g_{ij}(z,x) = g_{\left (0\right ) i j} + z^{2} g_{\left (2\right ) i j} + z ^{3}  g_{\left (3\right ) i j} + ...\,.
\label{g}
\end{equation}

\noindent
From this form of the metric, the following expressions are straightforwardly derived

\begin{align}
\sqrt{-h} &= \frac{\sqrt{g_{(0)}}}{z^{3}} + \mathcal{O} \left(z^{-1}\right)\,, \label{n1}\\
\left(h^{-1}\delta h\right)^{j}_{\ell} &= \left(g^{-1}_{(0)}\delta g_{(0)}\right)^{j}_{\ell} + \mathcal{O}(z^2) \,, \label{n2} \\
K^{i}_{j}\left(h\right) &= \frac{1}{\ell} \delta^{i}_{j} - \ell z^{2} S^{i}_{j}\left(g_{(0)}\right) + \mathcal{O}(z^{3}) \,, \label{n3}
\end{align}

\noindent
where $S^{i}_{j}$ is the Schouten tensor defined for the boundary metric $g_{(0)}$, i.e.,

\begin{equation}
S^{i}_{j} \left(g_{(0)}\right) = \mathcal{R}^{i}_{j} \left(g_{(0)}\right) - \frac{1}{4} \delta^{i}_{j} \mathcal{R}\left(g_{(0)}\right) \,. \label{n4}
\end{equation}

\noindent
In a similar fashion, one can compute the fall-off of the different components of the spacetime Weyl tensor.
Here, we just write down the ones which are of relevance for this holographic discussion

\begin{align}
W^{i z}_{jk} &= \mathcal{O}\left(z^{4}\right) \,, \label{n5}\\
W^{ik}_{jm} &= z^{2} \mathcal{W}^{ik}_{jl}\left(g_{\left(0\right)}\right) + \frac{3}{2} \frac{z^{3}}{\ell^{2}} g^{[i}_{\left(3\right)[j} \delta^{k]}_{m]} + \mathcal{O}\left(z^{4}\right) \,, \label{n7}
\end{align}

\noindent
where $\mathcal{W}^{ik}_{jm}$ correspond to the boundary Weyl tensor and the indices of $g_{(3)}$
are raised and lowered with the metric $g_{(0)}$.

Replacing all the above quantities in Eq.(\ref{varCG2}), we
realize that the first term and third terms in the integrand are of order $\mathcal{O} \left(z^{2}\right)$.
That implies that these terms do not contribute in the limit $z \rightarrow 0$.
In turn, the only nonvanishing contribution comes from the second term
in Eq.(\ref{varCG2}) as

\begin{equation} \label{Gueyl}
\delta I_{CG} = \frac{\ell}{16\pi G} \int \limits _{\partial M} d^{3}x \sqrt{g_{\left(0\right)}} \frac{3}{2\ell^{2}} g^{ij}_{\left(3\right)} \delta g_{\left(0\right)i j}\,,
\end{equation}

\noindent
expressed in terms of the holographic Einstein modes.

One can take a few steps back in the expansion of the boundary quantities and appropriately covariantize the
last result, in order to express it in terms of the subtrace of the spacetime Weyl tensor

\begin{equation} \label{Gueyl2}
\delta I_{CG} = \frac{\ell}{16\pi G} \int \limits_{\partial M} d^{3}x \sqrt{-h}W^{j\ell}_{i\ell}\left(h^{-1}\delta h\right)^{i}_{j}.
\end{equation}

\noindent
Due to the fact that the Weyl tensor is traceless ($W^{j\mu}_{i\mu}$), its subtrace can be traded
off by the electric part of the Weyl tensor

\begin{equation}
W^{j\ell}_{i\ell}=-W^{j z}_{i z}\,.
\end{equation}

\noindent
As a consequence, the variation of the Conformal Gravity action is

\begin{equation}
\delta I_{CG} = -\frac{\ell}{16\pi G} \int \limits _{\partial M} d^{3}x \sqrt{-h} E^{j}_{i}\left(h^{-1}\delta h\right)^{i}_{j}\,,
\end{equation}

\noindent
for the Einstein modes of the theory. At the same time, this means that the definition of conserved quantities for
that sector of CG can be mapped to the notion of Conformal Mass in $4D$ \citep{Ashtekar:1984zz}.

\section{Conclusions}

In the present work, we have shown that, in Critical Gravity, the energy of any
Einstein solution vanishes identically. This proof does not make use of any particular
Einstein black hole, nor relies on charge formulas obtained from the linearization of
the field equations. In this respect, charge expression (\ref{total_charge}) provides the explicit realization
of a claim originally stated in Ref.\citep{Porrati:2011ku}.\newline
The holographic derivation in Section \ref{electric} confirms the fact that the boundary stress tensor
for the total action (\ref{CritG_actionA}) is zero, in a similar way as in Ref. \citep{Johansson:2012fs}.\newline
When one goes beyond Einstein spaces, the expression (\ref{total_charge}) is able to
capture the effects due to the presence of higher-derivative terms in the curvature.
Indeed, as it was shown in Ref.\citep{Anastasiou:2016jix}, only the non-Einstein modes of
the Weyl tensor survive in the surface term form the variation of the Critical Gravity action.
 As a matter of fact, the boundary contributions are expressible in terms
of the Bach tensor, what enormously simplify the computation of holographic correlation
functions at the critical point \citep{Anastasiou:2017mag}.\newline
Noether-Wald charges provides the black hole entropy in a given gravity theory, when evaluated
at the horizon $r=r_{h}$,

\begin{equation}
S = - 2 \int \limits_{\Sigma_{h}}\,dS_{\nu}E^{0 \nu}_{0 \alpha}\nabla^{\alpha}\xi^{0}.
\end{equation}

\noindent
As the condition in the Weyl tensor (\ref{AdS_curvature}) holds throughout the spacetime for Einstein solutions,
it is evident from the above formula that the entropy vanishes in Critical Gravity.
The addition of topological invariants to the four-dimensional AdS gravity action has led to
energy definitions which are finite \citep{Olea:2005gb,Araneda:2016iiy}, but also has provided insight
on the problem of holography for asymptotically AdS spaces in Einstein gravity \citep{Miskovic:2009bm}.
The result presented here indicates that the Gauss-Bonnet term also plays a role in the holographic description
of gravity beyond Einstein theory.

\section{Acknowledgments}

We thank O. Miskovic for helpful discussions. G.A. is a Universidad Andres Bello (UNAB) Ph.D. Scholarship holder, and his work is supported by Direcci\'{o}n General de Investigaci\'{o}n (DGI-UNAB). D.R.B. is a UNAB M.Sc. Scholarship holder. This work was funded in part by FONDECYT Grant No. 1170765; CONICYT Grant No. DPI 20140115 and UNAB Grant No. DI-1336-16/R.

\bibliographystyle{ieeetr}
\bibliography{references}

\end{document}